\documentclass{PoS}

\title{Spontaneous CP-violation in the Simplest Little Higgs Model}

\ShortTitle{Spontaneous CP-violation in the Simplest Little Higgs Model}

\author{\speaker{Ying-nan Mao}\\
        Physics Division, National Center for Theoretical Sciences, Hsinchu, Taiwan 300\\
        E-mail: \email{ynmao@cts.nthu.edu.tw}, \email{maoyn@ihep.ac.cn}}

\abstract{We proposed how spontaneous CP-violation can be generated in the simplest little Higgs model in this talk.
Comparing with the original paper, both formalism and phenomenology are updated. The model is still alive facing the
collider and electric dipole moment (EDM) measurements. Strictest constraint comes from LHC $Z'$ direct search, which leads to $f\gtrsim8~\textrm{TeV}$. 
Higgs measurements also set strict constraint on the scalar mixing angle, if the Higgs rare decay channel is open. EDM measurements 
still set weak constraints for this model, even after the recent ACME updated measurement on electron's EDM. In this talk, we also 
discussed the test of CP-violation in the scalar sector, through the interactions between scalars and gauge bosons.}

\FullConference{The 39th International Conference on High Energy Physics (ICHEP2018)\\
		4-11 July, 2018\\
		Seoul, Korea}

\begin{document}

\section{Introduction}
Little Higgs mechanism \cite{LH} was proposed to solve the ``little hierarchy'' problem. As an example, the simplest little Higgs
(SLH) model \cite{SLH} contains the minimal extended scalar sector. In the standard model (SM), complex CKM matrix is the only source
of CP-violation. It is successful to explain all discovered CP-violation effects \cite{PDG}. However, there may be other CP-violation
sources, especially in the scalar sector. The speaker had proposed the possibility of spontaneous CP-violation in the SLH model
\cite{SCPVSLH} with the new SLH formalism \cite{Newform}, as an example of the connection between CP-violation and little Higgs
mechanism. In this talk, we introduce its realization together with the updated formalism and phenomenology.

\section{Formalism}
In the SLH model, a global symmetry breaking $[\textrm{SU}(3)\times\textrm{U}(1)]^2\rightarrow[\textrm{SU}(2)\times\textrm{U}(1)]^2$
happens at a scale $f\gg v=246~\textrm{GeV}$. The gauge group is also extended to $\textrm{SU}(3)\times\textrm{U}(1)$, which means there
are five extra heavy gauge bosons at scale $f$. Two scalar triplets are nonlinear realized as \cite{SLH2}
\begin{equation}
\Phi_1=\textrm{e}^{\textrm{i}\frac{\Theta'}{f}}\textrm{e}^{\textrm{i}\frac{t_{\beta}\Theta}{f}}\left(0,0,fc_{\beta}\right)^T,\quad
\Phi_2=\textrm{e}^{\textrm{i}\frac{\Theta'}{f}}\textrm{e}^{-\textrm{i}\frac{\Theta}{ft_{\beta}}}\left(0,0,fs_{\beta}\right)^T.
\end{equation}
We denote $s_{\alpha}\equiv\sin\alpha,c_{\alpha}\equiv\cos\alpha,t_{\alpha}\equiv\tan\alpha$ as usual. The matrix fields are
\begin{equation}
\Theta\equiv\frac{\eta}{\sqrt{2}}+\left(\begin{array}{cc}\mathbf{0}_{2\times2}&\phi\\ \phi^{\dag}&0\end{array}\right),\quad
\Theta'\equiv\frac{\zeta}{\sqrt{2}}+\left(\begin{array}{cc}\mathbf{0}_{2\times2}&\varphi\\ \varphi^{\dag}&0\end{array}\right).
\end{equation}
SM singlets $\eta,\zeta$ and doublets $\phi\equiv((v+h-\textrm{i}\chi)/\sqrt{2},G^-)^T,\varphi\equiv((\sigma-\textrm{i}\omega)/\sqrt{2},x^-)^T$
are all pseudo-Goldstone bosons, but only two of which are physical. $\phi$ is the usual Higgs doublet and thus $h$ is the physical Higgs boson.
We should also perform further canonically-normalization \cite{Newform} in the CP-odd scalar sector, and after which another physical field is
proportional to $\eta$, as expected.

In fermion sector, all doublets should also be enlarged to triplets. For example, for the third generation quark, we need another SM singlet
heavy quark $T$ as the partner of $t$. The Yukawa interaction is constructed based on the ``anomaly-free embedding'' \cite{SLH2,AF}. Electro-weak
symmetry breaking (EWSB) is mainly induced by quantum correction. To one-loop level, the scalar potential which may induce CP-violation in this model
is \cite{SCPVSLH,Newform,SLH2,CW,SLH3}
\begin{equation}
V=\left[-\mu^2\Phi_1^{\dag}\Phi_2+\epsilon\left(\Phi_1^{\dag}\Phi_2\right)^2+\textrm{H.c.}\right]+\lambda\left|\Phi_1^{\dag}\Phi_2\right|^2+
\left[\Delta_A+A\left(\ln\frac{v^2}{2\phi^{\dag}\phi}-\frac{1}{2}\right)\right].
\end{equation}
The coefficients in the last term are \cite{Newform}
\begin{equation}
\Delta_A=\frac{3}{16\pi^2}\left(\lambda_t^4\ln\frac{m^2_T}{m^2_t}-\frac{g^4}{8}\ln\frac{m^2_X}{m^2_W}-\frac{g^4}{16c^4_W}\ln\frac{m^2_{Z'}}{m^2_Z}\right),\quad
A=\frac{3}{16\pi^2}\left(\lambda^4_t-\frac{g^4}{8}-\frac{g^4}{16c^4_W}\right),
\end{equation}
with $\lambda_t\equiv\sqrt{2}m_t/v$. $X$ and $Z'$ are charged and neutral heavy gauge bosons respectively. Here we used the continuum effective
field theory (CEFT) framework \cite{CEFT} in which the dependence on the UV-cutoff $\Lambda$ is canceled during renormalization, as shown in \cite{Newform}. Define $\kappa\equiv v/f$ and $\alpha\equiv\sqrt{2}\kappa/s_{2\beta}$, when both $\mu,\epsilon\neq0$ and $\mu^2<|2\epsilon f^2s_{2\beta}c_{\alpha}|$, $\eta$
would acquire a nonzero VEV $v_{\eta}=f\xi s_{2\beta}/\sqrt{2}$ where $\xi\equiv\arccos\left(\frac{\mu^2}{2\epsilon f^2s_{2\beta}c_{\alpha}}\right)$. That
means spontaneous CP-violation occurs. From the potential, we have $\lambda=2\epsilon+(\Delta_A-A)\kappa^2(\alpha/s_{\alpha})$.

Both scalars become CP-mixing states, we denote them as $h_{1,2}$, the mass mixing has the form
\begin{equation}
h_1\equiv c_{\theta}h-s_{\theta}\eta,\quad h_2\equiv s_{\theta}h+c_{\theta}\eta,
\end{equation}
where the mixing angle satisfies $t_{2\theta}=\frac{4\epsilon f^2s_{2\xi}s_{\alpha}}{4\epsilon f^2s^2_{\xi}-M^2_{hh}}$. The masses are
\begin{equation}
m_{1,2}^2=\frac{M^2_{hh}+4\epsilon f^2s^2_{\xi}}{2}\pm\left(\frac{M^2_{hh}-4\epsilon f^2s^2_{\xi}}{2}c_{2\theta}-2\epsilon f^2s_{2\xi}s_{\alpha}s_{2\theta}\right),
\end{equation}
where $M^2_{hh}$ is the mass matrix element $-\left.\frac{\partial^2V}{\partial h^2}\right|_{h=\eta=0}=4\epsilon f^2c^2_{\xi}s^2_{\alpha}
+v^2\left[\left(3-\frac{2\alpha}{t_{2\alpha}}\right)\Delta_A-\left(5-\frac{2\alpha}{t_{2\alpha}}\right)A\right]=m_1^2c^2_{\theta}+
m^2_2s^2_{\theta}\simeq m^2_1$. Here we chose $|\theta|\ll1$ thus $h_1$ is the SM-like scalar.

The $h_1h_2h_2$ trilinear interaction can be parameterized as $\mathcal{L}\supset-\lambda_Ifh_1h_2^2/2$, in which
\begin{equation}
\lambda_I=c_{\theta}s^2_{\theta}\lambda_0+s_{\theta}(2-3s^2_{\theta})\frac{\sqrt{2}\epsilon s_{2\xi}(3c_{2\alpha}-1)}{s_{2\beta}c_{\alpha}}+
c_{\theta}(1-3s^2_{\theta})\frac{2\sqrt{2}\epsilon t_{\alpha}(3c_{2\xi}-1)}{s_{2\beta}}-c^2_{\theta}s_{\theta}\frac{6\sqrt{2}\epsilon s_{2\xi}}{c_{\alpha}s_{2\beta}},
\end{equation}
with $\lambda_0\equiv2(3\Delta_A-8A)\kappa+8(\Delta_A-A)\kappa^3/s^2_{2\beta}+6\sqrt{2}\epsilon c^2_{\xi}s_{2\alpha}/s_{2\beta}$. When $m_2<m_1/2$,
such a vertex can lead to $h_1\rightarrow2h_2$ rare decay, which will face strict constraint from Higgs measurements.

\section{Phenomenology}
Facing many experimental constraints, the model is still alive. First, through direct search of $Z'$ at LHC \cite{Zprime}, we obtained
a $95\%$ C.L. lower limit $f\gtrsim8~\textrm{TeV}$. Its upper limit is about $f\lesssim85~\textrm{TeV}$ \cite{Newform}, which is the same as the
CP-conserving case based on Goldstone scattering unitarity analysis \cite{Newform}. The bounds on $s_{\theta}$ mainly come from the Higgs signal strengths fit, LEP and LHC \cite{stbound} direct searches. When $m_2\lesssim m_1/2$, strict constraint comes from the bound for $h_1\rightarrow2h_2$ rare decay channel. For light $m_2$ with a mass around $(15-60)~\textrm{GeV}$, we have $|s_{\theta}|\lesssim(0.03-0.15)$; while for large $m_2(\gtrsim m_1/2)$, we have $s_{\theta}\lesssim(0.15-0.3)$ \cite{SCPVSLH}. With the updated CEFT formalism, we also obtained $12~\textrm{TeV}\lesssim m_T\lesssim24~\textrm{TeV}$ in this model. As a model with extra CP-violation source, there must be constraints from electric dipole moments (EDM) \cite{EDM}. In this model, this constraint is not very strict, even after the recent updated measurement of electron EDM \cite{EDM}. For $|s_{\theta}|\sim(0.1-0.2)$ and $f\gtrsim8~\textrm{TeV}$, it is allowed for $m_2\sim(20-500)~\textrm{GeV}$. Constraints from neutron and heavy atoms' EDM set weaker constraints comparing with electron EDM.

To test the CP-violation in the scalar sector, we just need to confirm both $h_2VV$ and $h_1h_2V$ vertices together (where $V$ denotes a massive gauge boson, and $h_1VV$ vertices have already been confirmed at LHC), based on the CP properties' analysis \cite{TestCPV}. For $h_2VV$ vertices, we can test it at a Higgs or $Z$ factory for light $h_2$ scenario, or at a hadron collider for heavy $h_2$ scenario. For $h_1h_2V$ vertices, we need the help of $Z'$ since the $Zh_1h_2$ vertex is suppressed by $\kappa^3$ \cite{Newform}, while the $Z'h_1h_2$ vertex is only suppressed by $\kappa$. This is one of the motivations for future colliders.

\section{Summary}
In this talk, we discussed how spontaneous CP-violation can be generated in the simplest little Higgs model. Comparing with the original paper, we updated both its formalism and phenomenology. The formalism is based on CEFT in the scalar potential, together with a canonically-normalization in the scalar kinetic part. The model is still allowed by data, facing the constraints from colliders and EDM. The most strict constraint for the model is $f\gtrsim8~\textrm{TeV}$, which comes from LHC direct $Z'$ search. For $m_2<m_1/2$, $s_{\theta}$ also face strict constraint from $h_1\rightarrow2h_2$ rare decay. We also discussed the test of CP-violation in the scalar sector. The idea is to confirm both $h_2VV$ and $h_1h_2V$ type vertices at future colliders.

\section*{Acknowledgement}
This work was partly supported by the China Postdoctoral Science Foundation (Grant No. 2017M610992). The speaker was also partly supported by the ICHEP2018 Travel Grant.

\end{document}